\documentclass{article}

\usepackage[nonatbib, final]{nips_2016}

\usepackage[utf8]{inputenc} 
\usepackage[T1]{fontenc}    
\usepackage{hyperref}       
\usepackage{url}            
\usepackage{booktabs}       
\usepackage{amsfonts}       
\usepackage{nicefrac}       
\usepackage{microtype}      
\usepackage{amsmath}
\usepackage{graphicx}
\usepackage{color}
\usepackage{algorithm}
\usepackage{algorithmic}
\usepackage{wrapfig}

\DeclareMathOperator*{\Unif}{Unif}
\DeclareMathOperator*{\Mean}{Mean}

\newcommand{\SKOsuggest}{\ensuremath{\text{suggestion\_{}from\_{}SKO}}}

\newcommand{\jinit}{{j_\text{init}}}
\newcommand{\jmax}{{j_{\text{max}}}}
\newcommand{\ktest}{{k_{\text{test}}}}
\newcommand{\nmax}{{n_{\text{max}}}}
\newcommand{\nmin}{{n_{\text{min}}}}
\newcommand{\navg}{{n_\text{avg}}}

\newcommand{\pp}{{\mathbf{p}}}

\newcommand{\ttheta}{{\boldsymbol{\theta}}}
\newcommand{\vv}{{\mathbf{v}}}

\newcommand{\xxi}{{\boldsymbol{\xi}}}

\newcommand{\cR}{\mathcal{R}}

\newcommand{\algref}[1]{\hyperref[#1]{Algorithm~\ref{#1}}}
\newcommand{\figref}[1]{\hyperref[#1]{Figure~\ref{#1}}}
\newcommand{\tabref}[1]{\hyperref[#1]{Table~\ref{#1}}}
\newcommand{\secref}[1]{\hyperref[#1]{Section~\ref{#1}}}

\title{Preemptive Termination of Suggestions during Sequential Kriging Optimization of a Brain Activity Reconstruction Simulation}

\author{
	Michael McCourt, Ian Dewancker \\
	SigOpt\\
	244 Kearny St.\\
	San Francisco, CA 94122 \\
	\texttt{\{mccourt, ian\}@sigopt.com} \\
	\And
	Salvatore Ganci \\
	Worldenergy \\
	Via Cantonale\\
	6562 Soazza GR, Switzerland\\
	\texttt{s.ganci@wdenergy.com} \\
}

\begin{document}

\maketitle

\vspace{-.1in}
\begin{abstract}
	Reconstructing brain activity through electroencephalography requires a boundary value problem (BVP) solver to take a proposed
	distribution of current dipoles within the brain and compute the resulting electrostatic potential on the scalp.
	This article proposes the use of sequential kriging optimization
	to identify different optimal BVP solver parameters for dipoles located in isolated sections of the
	brain by considering the cumulative impact of randomly oriented dipoles within a chosen isolated section.
	We attempt preemptive termination of parametrizations suggested
	during the sequential kriging optimization which, given the results to that point,
	seem unlikely to produce high quality solutions.  Numerical
	experiments on a simplification of the full geometry for which an approximate solution is available show a benefit
	from this preemptive termination.
\end{abstract}

\section{Introduction\label{sec:introduction}}

Electroencephalography (EEG) is a non-invasive tool for localizing neural sources and reconstructing brain
activity using measurements on the surface of a patient's scalp \cite{NiedermeyerDasilve12};
its practical applications involve both clinical diagnoses and neurophysical research.  Maxwell's equations explain
the mechanism through which knowledge of the location and orientation of current dipoles
(which model beams of active neurons) within the brain can be
used to predict the resulting electrostatic potential on the scalp \cite{NunezSrinivasan06}.  In practice, then,
these electrostatic potentials can be measured on the scalp of a patient under some defined stimulus and an
inverse problem can be solved to reconstruct the current distribution within the brain \cite{GrechCassarEtAl08}.

As is the case with most inverse problems, the efficiency of the solution mechanism is strongly dependent on the
quality of the \emph{forward solver}; for this problem, the forward solver is the boundary value problem (BVP)
solver which solves Maxwell's equation for a proposed dipole or set of dipoles which produce the boundary
conditions.  Possible solvers for this forward problem include
the finite element method \cite{WoltersKostlerEtAl08}, the more popular
boundary element method (BEM) \cite{KybicClercEtAl05} and, our preferred strategy,
the method of fundamental solutions (MFS) \cite{AlaFasshauerEtAl15}.

The MFS may be preferred to FEM/BEM for its meshfree nature but it also comes with complications,
primarily in the form of free parameters which may provide good or bad numerical accuracy depending on how they
are chosen.
Some literature describes special circumstances for which optimal MFS parameter choices are known \cite{Katsurada90},
but for most problems there are only heuristic parametrization strategies for individual applications \cite{ChenKarageorghisEtAl16}.

\secref{sec:mfsdefinition} details the MFS computational situation and the role that the various parameters play.
In \secref{sec:methodology} we define a metric which measures the quality of given parameter values;
we also propose a small modification to the standard sequential kriging optimization (SKO) workflow
to minimize computation on underperforming parameter suggestions.
\secref{sec:experiments} contains numerical experiments on a simplified version of the problem which
show the viability of Bayesian optimization for parameter tuning in this setting.

\subsection{The MFS for the Coupled BVP\label{sec:mfsdefinition}}
The full formulation of this EEG problem using the MFS is rather involved and is discussed in full detail in
\cite{AlaFasshauerEtAl15}, with \cite{AlaFasshauerEtAl15b} also providing the structure for the nearby
magnetoencephalography problem.  We limit this discussion to identifying the role of the free
parameters in defining the MFS solution.

\begin{figure}[ht]
	\centering
	\includegraphics[width=.9\textwidth]{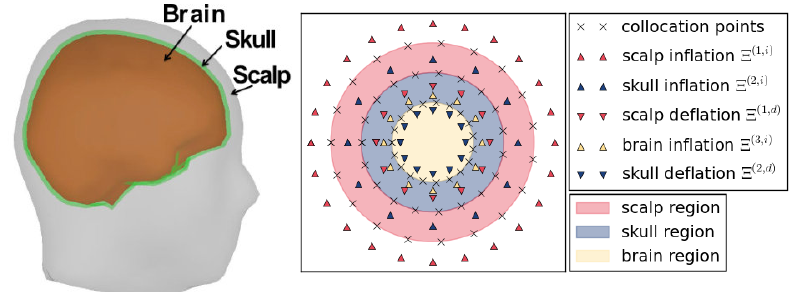}
	\caption{\label{fig:mfsdomain}
		\textit{left}: Depiction of the components of the reconstruction problem.
		\textit{right}: Distribution of MFS points on a more easily readable simplification of the domain.
	}
\end{figure}

The MFS is popular for certain boundary value problems because it converts a boundary
value problem over a volume into an approximation problem on the boundary \cite{ChenKarageorghisEtAl08}.
In this situation, there are three boundary value problems, one each within the brain/skull/scalp, which must be
solved in a coupled fashion to define the electrostatic potential on a patient's scalp given a current dipole or set of dipoles within
the brain.  A separate solution must be defined on each of these domains,
\begin{align*}
	u^{(r)}(\pp) = \sum_{\xxi_j\in\Xi^{(r, \text{i})}} c_j^{(r, \text{i})} K(\pp, \xxi_j) + \sum_{\xxi_j\in\Xi^{(r, \text{d})}} c_j^{(r, \text{d})} K(\pp, \xxi_j),
\end{align*}
for $r=1, 2, 3$ corresponding to the scalp, skull and brain\footnote{
	The brain has no deflation region, thus $\Xi^{(3, \text{d})}$ is empty.},
respectively.
and each of these solutions has free parameters,
Here, $K(\pp, \xxi) = 1 / \|\pp - \xxi\|$ is the Green's kernel for the three dimensional Laplacian on an unbounded domain
\cite{FairweatherKarageorghis98}.
Our free parameters appear in the construction of the so-called \emph{fictitious boundary} on which
the sets of kernel centers $\Xi$ are defined; we use $\Xi^{(1, \text{i})}$ to denote the kernel centers on the
inflated fictitious boundary associated with 
the scalp solution and $\Xi^{(1, \text{d})}$ to define the kernel centers on the scalp's deflated fictitious boundary\footnote{
	Because some of the collocation conditions appear as part of coupling between domains (brain to skull and skull to scalp),
	this might be more accurately called a \emph{fictitious interface} for those components.
	We retain the term fictitious boundary to match existing literature despite the fact that only the scalp has a boundary condition.
}.
The allocation and placement of source points has been a complication for MFS methods
since their inception and was discussed at length in a recent survey \cite{ChenKarageorghisEtAl16}.
\figref{fig:mfsdomain} depicts the computational setup for this problem.

Following the problem definition in \cite{AlaFasshauerEtAl15}, each fictitious boundary is defined with an
inflation or deflation parameter, and the size of the set of kernel centers $\Xi$ associated with that fictitious
boundary, $|\Xi|$, must also be chosen.\footnote{
	It is also an interesting question if even the individual location of each kernel center could be
	optimally chosen, but we assume, for now, that centers are as uniformly distributed as possible
	on their fictitious boundary.  There is also some theoretical guidance regarding acceptable kernel center
	locations \cite{BarnettBetcke08}.}
Thus, in this formulation, there are as many as 10 free parameters defining the solution to this coupled BVP,
two for each fictitious boundary/interface.  However, because there is often a desire to balance the number of kernel
centers and collocation points, we fix values for the number of centers $|\Xi|$ and instead attempt
to only adjust the inflation/deflation of the 5 fictitious boundaries.  We hope to use future work to simultaneously
select both aspects of the BVP solver.

\section{Sequential Kriging Optimization for Tuning MFS EEG Solvers\label{sec:methodology}}
Sequential kriging optimization
\cite{Huang06} as well as related terms and methods such as hyperparameter selection
\cite{BergstraBardenetEtAl11}, sequential model-based algorithm configuration \cite{HutterHoosEtAl11}
or the ultimately general Bayesian optimization \cite{SnoekLarochelleEtAl12}, provide strategies for
efficiently optimizing black-box functions.
To more clearly state our black-box function, we assume only
\emph{a single} dipole $\pp'$ exists within the brain; this distinction would need to be lifted
to apply this method in the full spectrum of possible EEG reconstructions.

The black box function we define here is based on the goal of identifying optimal free parameters for an
MFS solution as the forward solver within an inverse solver so as to make the inverse solver as efficient and
accurate as possible. We define the quality of a set of parameters $\ttheta$ for
\emph{a single dipole} $\pp'$ as the difference between the MFS and true solutions for that $\pp'$
at a set of test points on the scalp\footnote{
	When the true solution is unavailable
	(as is almost certainly the case)
	we substitute an expensive BEM solution computed with a very high number of elements so as to be
	substantially more accurate than the MFS solution.
	In future work, we hope to define quality in a cross-validation style
	computation within the inverse solver where the true solution is represented by a validation set drawn
	from the actual observations.}.
We compress this difference into a scalar with
\begin{align}
	\label{eq:qualitymetric}
	Q(\ttheta; \pp') = -\log\left(\sum_{k=1}^{\ktest} \left|u_\text{MFS}^{(1)}(\pp_k) - u_\text{true}^{(1)}(\pp_k) \right|^2 \;\Big/\;
		\sum_{k=1}^{\ktest} \left|u_\text{true}^{(1)}(\pp_k) \right|^2\right),
\end{align}
where $\ktest$ represents the number of $\pp_k$ test points placed on the scalp.
The dipole $\pp'$ appears implicitly in the definition of the $c$ terms in the $u_\text{MFS}$ solution as well
as in the ``true'' solution.

Dipoles in different brain regions will likely require different
inflation/deflation of the fictitious boundary to perform optimally.  We propose that, within a given region $\Omega'$ of
the brain, dipoles be selected with random location and orientation\footnote{
	The appropriate random selection may be complicated based on the neurophysiology of that
	region.  We leave that discussion for future research and simply choose dipoles of magnitude unity
	uniformly distributed within $\Omega'$ with uniform distribution over orientations.}
and then we study the mean performance over those dipoles.
If we abuse notation slightly and define the uniform sampling of dipole locations
and orientations within $\Omega'$ as $\pp'\sim\Unif(\Omega')$, we can denote the sample mean over $\navg$
observations as $\bar{Q}(\ttheta)$; \emph{this sample average} is the black-box function to be maximized during
the SKO.  We also use the sample variance in the construction of the
kriging model \cite{RasmussenWilliams06}.

\begin{wrapfigure}[17]{l}{.45\textwidth}
	\vspace{-.75cm}
	\begin{minipage}{.45\textwidth}
		\begin{algorithm}[H]
			\caption{Preemptive EEG Termination\label{alg:earlyeegterminationn}}
			\begin{algorithmic}
				\STATE {\bfseries Input:}  $\Omega', \jmax, \navg, \nmin, \jinit$
				\STATE $ j \gets 0; \cR \gets \{\}; $
				\WHILE{$j < \jmax$}
				\STATE $\ttheta \gets \SKOsuggest(\cR)$
				\STATE $\vv \gets \{\}$
				\FOR{$n \gets 0$ \TO $\navg$}
				\STATE $\pp'\sim\Unif(\Omega')$
				\STATE $\vv \gets \vv \cup Q(\ttheta; \pp')$
				\STATE $j \gets j + 1$
				\IF{$n > \nmin$ \AND $j > \jinit$}
				\STATE \textbf{break if  $\Mean(\vv)<\Mean(\cR)$}
				\ENDIF
				\ENDFOR
				\STATE $\cR \gets \cR \cup \{\ttheta, \vv\}$
				\ENDWHILE
			\end{algorithmic}
		\end{algorithm}
	\end{minipage}
\end{wrapfigure}

In initial SKO experiments, we recognized that for
$\pp'\sim\Unif(\Omega')$ most $\ttheta$ yielded similar distribution shapes of $Q(\ttheta; \pp')$ values differing
primarily by their mean.  This led us to believe that, since the impact of $\ttheta$ was primarily a translation of
the $\bar{Q}$ distribution, we could predict with fewer than
$\navg$ values of $\pp'\sim\Unif(\Omega')$ whether a given $\ttheta$ could perform optimally.

We implemented a heuristic strategy which is on display in \algref{alg:earlyeegterminationn}.
This algorithm requires choosing preemption parameters: $\jmax$ is the total number of MFS/true
solutions to be computed over the course of the optimization, $\navg$ is the maximum number of $\pp'$ choices
to consider before returning $\bar{Q}(\ttheta)$, $\nmin$ is the minimum number of $\pp'$ choices to consider
before allowing preemptive termination, and $\jinit$ is the number of initial $Q$ evaluations for which preemption
is forbidden so as to form a baseline of $Q$ values.
We write $\Mean(\cR)$ to denote the mean of the scores of \emph{all} the $\pp'$ considered thus far when pooled
into a single sample, i.e., if $\cR=\{\{\ttheta_1, \{1, 2, 3\}\}, \{\ttheta_2, \{4, 5, 6\}\}\}$ then $\Mean(\cR)=3.5$.
Any partially completed $\ttheta$ values still report their $\bar{Q}$ and sample variance values to the kriging model.

This concept of a variable amount of work depending on the perceived quality of a given suggestion appears
under various names in the
literature such as freeze-thaw Bayesian optimization \cite{SwerskySnoekEtAl14}, hyperband \cite{LiJamiesonEtAl16},
or predictive termination \cite{DomhanSpringenbergEtAl15}.
Similarly to our strategy, the F-Race Algorithm \cite{BirattariYuanEtAl10} (which has an open source implementation
\cite{LopezEtAl16}) provides a mechanism for allowing multiple parallel suggestions and a statistical analysis for discarding
underperforming suggested $\ttheta$.  Similar racing and adaptive capping methods are also available in the
open source software ParamILS \cite{HutterHoosEtAl09}.
The literature on multi-armed or infinitely-armed bandits may also provide a more rigorous mechanism for pausing
and revisiting a range of $\ttheta$ suggestions in a regret-based framework, rather than terminating suggestions
as described in \algref{alg:earlyeegterminationn} \cite{Auer02}.
Of immediate interest to our SKO framework is the use of covariance kernels which contain components modeling the
quality of the approximation $\bar{Q}$ as a function of $n$ (and, perhaps, a bootstrap estimate of the variance \cite{Efron81})
which is a topic that has appeared in some uncertainty quantification research \cite{PichenyGinsbourger13}.
While our heuristic is simple to state and implement, it will be improved upon and augmented in future work using this rich breadth
of available ideas present in the optimization and model configuration communities.

\section{Numerical Experiments\label{sec:experiments}}

To simplify initial testing and isolate potential sources of error/uncertainty, these numerical experiments
involve a less complicated geometry for which there exists an analytic solution \cite{Zhang95}:
three concentric spheres (of radii .087, .092 and .1~m
with conductivity .33, .0125 and .33~S/m respectively)
replace the physical brain/skull/scalp geometry we eventually hope to use.
Refer to \figref{fig:mfsdomain} to see how the source points are distributed in this domain.
Future work must involve the more realistic geometries explored in \cite{AlaFasshauerEtAl15}.

All dipole moments have magnitude 1 A$\cdot$m.
The SKO used an expected improvement acquisition function \cite{JonesSchonlauEtAl98};
any $\ttheta$ values that yielded a collocation matrix without full rank were treated as failures.
These experiments used values of $\ktest=1000$, $\jmax=800$, $\navg=30$, $\nmin=5$ and $\jinit=5\navg$;
for these parameters, the maximum number of non-failed $\ttheta$ values which can be considered (which only occurs in
the case where they produce increasingly low $\bar{Q}$) is $\jinit/\nmax + (\jmax-\jinit)/\nmin=135$ and the minimum number
to be considered (which only occurs for increasingly high $\bar{Q}$) is $\jinit/\nmax + (\jmax-\jinit)/\nmax\approx27$.
The parameter values suggested here are chosen arbitrarily, although $\nmin=5$ was provide some opportunity to
perform before truncation.
All point distributions on spheres are created with the spiral method described in \cite{SaffKuijlaars97}.
The number of collocation points on each of the three spheres was always fixed at 300,
with $|\Xi^{(1, d)}| = |\Xi^{(2, i)}| = |\Xi^{(2, d)}| = |\Xi^{(3, i)}| = 90$ and $|\Xi^{(1, i)}| = 180$; these choices were made
arbitrarily, and future work could include tuning these values along with the inflation/deflation parameters.
Test results are presented in \tabref{fig:optimresults}.

\begin{table}[ht]
	\centering
	\caption{\label{fig:optimresults}Median results over 30 tests from specific dipole distributions.
		\textsc{Preemptive} and \textsc{standard}
		refer to whether or not $\nmin<\navg$ or $\nmin=\navg$, respectively.  \textsc{Bonus}
		states the number of extra suggestions enjoyed by the preemptive strategy;
		for example, a \textsc{Bonus} 10 value
		states that 10 additional $\ttheta$ parameters were able to be tested because the preemptive truncation
		allowed for as few as $\nmin$ MFS solves to be conducted rather than the full $\navg$.  \textsc{Preemptive}
		results in \emph{italics} indicate a statistically significant difference with $p<.05$ from \textsc{Standard} using a Mann-Whitney $U$ test.
		}
	\begin{tabular}{c |ccc |ccc} \hline 
		\sc Index& \multicolumn{3}{|c}{\sc Sequential Kriging Optimiztion} & \multicolumn{3}{|c}{\sc Random Search} \\
		&\sc Standard&\sc Preemptive&\sc Bonus&\sc Standard&\sc Preemptive&\sc Bonus \\ \hline
		1 & 3.681&\textbf{\textit{3.845}}&16 &3.631&\textit{3.439}&27 \\
		2 & 1.494&\textbf{\textit{1.532}}&15 &1.333&1.383&30 \\
		3 & 5.090&\textbf{\textit{5.245}}&10 &4.672&\textit{4.905}&19 \\
		4 & 0.862&\textbf{\textit{0.978}}&28 &0.778&\textit{0.870}&34 \\
		5 & 0.219&\textbf{\textit{0.257}}&21 &0.146&\textit{0.195}&30 \\
		6 & 2.131&\textbf{\textit{2.209}}&21 &1.944&\textit{2.058}&31 \\ \hline
	\end{tabular}
\end{table}

At \verb+https://github.com/sigopt/sigopt-examples+ there is a description of the dipole regions
associated with each \textsc{index}.  As evidenced by the range of optimal $\bar{Q}$ values found,
the tests vary in difficulty; however, in each test there is a benefit to using SKO
over basic random search \cite{BergstaBengio12}, and a benefit to using the
preemption strategy outlined in \secref{sec:methodology}.

\small

\subsubsection*{Acknowledgments}
We would like to present our thanks to the organizers of this workshop, and especially to Roberto Calandra who has been
very helpful and supportive throughout this process.  We also would like to thank the anonymous referee whose
in depth comments and comprehensive references have greatly improved our presentation and have provided us with
numerous future paths for which to improve this research.

\bibliographystyle{plain}

\end{document}